\documentclass[twocolumn,showpacs,preprintnumbers,floatfix,amsmath,amssymb,superscriptaddress]{revtex4}

\usepackage{graphicx}
\usepackage{dcolumn}
\usepackage{bm}

\newcommand{\be}{\begin{equation}}
\newcommand{\ee}{\end{equation}}
\newcommand{\bn}{\begin{eqnarray}}
\newcommand{\en}{\end{eqnarray}}
\newcommand{\ba}{\begin{array}}
\newcommand{\ea}{\end{array}}
\newcommand{\bc}{\begin{center}}
\newcommand{\ec}{\end{center}}
\newcommand{\bml}{\begin{mathletters}}
\newcommand{\eml}{\end{mathletters}}

\renewcommand{\bbox}[1]{\bm{#1}}

\begin{document}

\preprint{ }

\title{Isospin mixing and the continuum coupling in weakly bound nuclei}

\author
{N. Michel}
\affiliation{CEA/DSM/IRFU/SPhN Saclay, F-91191 Gif-sur-Yvette, France}%
\affiliation{Department of Physics, Post Office Box 35 (YFL), University of Jyv{\"a}skyl{\"a}, FI-40014 Jyv{\"a}skyl{\"a}, Finland}%

\author{W. Nazarewicz}
\affiliation{
Department of Physics and Astronomy, University of Tennessee, Knoxville, Tennessee 37996, USA
}%
\affiliation{
Physics Division, Oak Ridge National Laboratory, Oak Ridge, Tennessee 37831, USA
}%
\affiliation{
Institute of Theoretical Physics, University of Warsaw, ul. Ho\.za 69,
PL-00-681 Warsaw, Poland }%

\affiliation{School of Engineering and Science,
University of the West of Scotland,
Paisley  PA1 2BE, United Kingdom} 

\author{M. P{\l}oszajczak}
\affiliation{
Grand Acc\'el\'erateur National d'Ions Lourds (GANIL), CEA/DSM - CNRS/IN2P3,
BP 55027, F-14076 Caen Cedex, France
}%

\date{\today}

\begin{abstract}
The isospin breaking effects due to the Coulomb interaction in weakly-bound nuclei are studied using the Gamow Shell Model, a complex-energy configuration interaction approach which simultaneously  takes into account  many-body correlations between valence nucleons and  continuum effects. We investigate 
the near-threshold behavior of  one-nucleon spectroscopic factors and the structure of wave functions along an isomultiplet.  Illustrative calculations are carried out for the $T$=1  isobaric  triplet. By using a shell-model Hamiltonian consisting of an isoscalar nuclear interaction and the Coulomb term, we demonstrate that for weakly bound or unbound systems the structure of isobaric analog states varies within the isotriplet and impacts the energy dependence of spectroscopic factors. We discuss the partial dynamical isospin symmetry present in   isospin-stretched systems,  in spite of  the Coulomb interaction that gives rise to  large mirror symmetry breaking effects.
\end{abstract}


\bigskip

\maketitle

\section{Introduction}

The charge independence of   nuclear force gives rise to isospin
symmetry \cite{[Hei32],[Wig37]} and the formalism of isotopic spin has
proven to be a very powerful concept in nuclear physics \cite{Wil69}.
While useful, isospin symmetry is not perfectly conserved. On the hadronic
level, isospin is  weakly violated due to the difference in the masses of the
up and down quarks \cite{Mil06,Mei08,Mac01}. The main source of
isospin breaking in atomic nuclei lies, however, in the electromagnetic interaction \cite{[Ber72]}.  

The  members of a nuclear  isomultiplet, in particular mirror nuclei, provide a unique playground for studying  isospin physics.
The  invariance under rotations in isospin space implies that energies of excited states in an  isomultiplet should be identical;
the  deviations are usually attributed to the Coulomb force \cite{Wil69,Nol69,Len06,War06,Ben07}. However, for nuclear states close to, or above the reaction thresholds, the isospin breaking can be modified by the coupling to the particle continuum.  
Here, a spectacular example  is the Thomas-Ehrman (TE) effect \cite{Ehr51,Tho52,Lan58} that occurs when  one of
the mirror states is unstable against particle emission due to a large asymmetry between  proton and neutron emission thresholds. 
The resulting TE energy shifts  strongly depend on the angular momentum content of the nuclear state and can be fairly large for low partial waves \cite{Com88,Wap02}.

The TE effect   has also  a direct consequence for the structure of mirror wave functions \cite{Aue00,Gri02,Tim08,Tim08a}. Indeed, for near-threshold states, the configuration mixing involving scattering states  strongly depends on (i) positions of particle emission thresholds in mirror systems (the binding energy effect) \cite{Oko08}, and (ii)  different asymptotic behavior of  neutron and proton wave functions. The latter  leads to the universal behavior of cross sections \cite{Wigner,Breit} and spectroscopic factors (SFs) \cite{Mic07,Mic07a}  in the vicinity of  a reaction threshold. 

Recently, SFs and asymptotic normalization coefficients have been discussed in mirror  systems within  cluster approaches \cite{Tim08,Tim08a}, and strong mirror symmetry-breaking in mirror SFs has been predicted. The main focus of this work is on the isospin  mixing and mirror symmetry breaking in the isobaric analog states (IAS) of light nuclei. We show how the different asymptotic behavior within an isomultiplet and the  isospin-nonconserving (INC) Coulomb interaction  impact wave functions of IASs and resulting SFs. Our theoretical framework is the complex-energy continuum shell model, the Gamow Shell Model (GSM) \cite{Mic02,Bet02,Mic04,Mic09}. GSM is a configuration-interaction approach  with a single-particle (s.p.) basis given by the Berggren ensemble \cite{Berggren1} which consists of Gamow (bound and  resonance) states and the  non-resonant scattering continuum.

This paper is organized as follows. Section~\ref{model} presents the details of the GSM calculations, with a particular focus  on the treatment of the Coulomb potential and the recoil term. SFs in IASs are discussed in Sec.~\ref{spectfac}. Therein, we study the dependence of SFs on the position of one- and two-particle thresholds.  Our calculations  are performed for prototypical $T=1$ isotriplet consisting of $J^\pi=0^+$ and 2$^+$ IASs in 
`$^6$He', $^6$Li,  and `$^6$Be'. To remove the binding energy effect, we assume {\em identical} 1n/1p emission thresholds. In this way, we isolate the effect of the continuum coupling on  isospin mixing, and study it in the vicinity of proton and neutron drip lines. The results for  $^6$He, $^6$Li, and $^6$Be are discussed in Sec.~\ref{HeLiBe} by considering experimental and predicted one-particle thresholds. We point out that the conservation of isospin  in the low-lying states of $^6$Be can be explained in terms of partial dynamical isospin symmetry present in the GSM wave functions of this isospin-aligned system. Finally, the conclusions are contained in Sec.~\ref{conclusions}.

\section{The model}\label{model}

The GSM Hamiltonian is diagonalized in the many-body Slater determinants spanned upon the Berggren s.p.~basis. The many-body resonant states of GSM obey the  generalized variational principle \cite{Rot09}; 
they are obtained using the generalized Davidson procedure that has been developed analogously to the generalized Lanczos procedure in the context of GSM (see Refs.~\cite{Mic02,Mic09} for details). 

We assume in the following that the nucleus can be described as a system of $n_\pi$ valence protons or $n_\nu$ valence neutrons evolving around a closed core. 
Since our discussion concerns the  isobaric triplet $^6$He-$^6$Li-$^6$Be, 
we take  $^4$He as a core. Consequently, the nuclei $^5$He and $^5$Li can be considered as one-particle systems, and $^6$He, $^6$Li, and $^6$Be as two-particle systems.
In calculations involving $^5$He and $^6$He, the s.p.~basis is generated
by a Woods-Saxon (WS) potential with the radius $R_0$=2\,fm, 
diffuseness $d$=0.65\,fm, and  spin-orbit strength $V_{\rm so}$=7.5\,MeV. The depth of the central potential $V_0$ has been  varied to move the binding energy of a one-neutron system, $^{5}$He (i.e., the one-neutron threshold).  For $V_0$=47\,MeV (the ``$^{5}$He" parameter set), this potential reproduces 
energies and widths of experimental  $3/2_1^-$ and $1/2_1^-$   resonances  in $^5$He. 

The GSM results should be  free from spurious center-of-mass (CM) motion. 
To cope with this problem in our GSM approach, we adopt a system of intrinsic nucleon-core coordinates inspired by the Cluster Orbital Shell Model (COSM) \cite{Suz88,Suz90}.
In the COSM coordinates, the translationally-invariant GSM Hamiltonian  can be written as:
\begin{equation}
\label{GSMCOSM}
\textit{H} = \sum_{i=1}^{n_\pi+n_\nu}\left [ \frac{\bbox{p}_i^2}{2\mu} + U_{i} \right] + \sum_{i<j}^{n_\pi+n_\nu} \left[ V_{ij} + \frac{1}{A_{c}} \bbox{p}_i\bbox{p}_j \right],
\end{equation}
where  $\mu$ is the reduced mass of the nucleon+core system, $U_i$ is the  one-body WS potential representing the field of the core, $V_{ij}$ is the two-body residual interaction between valence nucleons, 
and the two-body  term $A_c^{-1}\bbox{p}_i\bbox{p}_j$, with  $A_c$ being the mass of the core, takes into account the recoil of the active nucleons.

The  modified finite-range surface Gaussian interaction (MSGI)   used in this study  is a variant of the finite-range surface Gaussian interaction (SGI) \cite{Mic04}. 
In order to discuss the motivation behind MSGI, we begin with the definition of the two-body residual interaction SGI:
\begin{eqnarray}
&&V_{J,T}^{SGI} (\bbox{r_1},\bbox{r_2}) \nonumber \\ &=& V_0(J,T) \exp \left[ - \left( \frac{\bbox{r_1} - \bbox{r_2}}{\mu_I} \right)^2 \right] \delta(r_1 + r_2 - 2 R_0) \nonumber \\
&=& V_0(J,T) \sum_{\ell=0}^{+ \infty}  \exp \left( - \frac{r_1^2 + r_2^2}{\mu_I^2} \right) \delta(r_1 + r_2 - 2 R_0) \nonumber \\ 
&\times&  i^\ell  (2 \ell + 1) ~ j_\ell \left( \frac{2 r_1 r_2}{i \mu_I^2} \right) \bbox{Y}_\ell(\hat{r}_1) \cdot \bbox{Y}_\ell(\hat{r}_2) \label{SGI_inter},
\end{eqnarray}
where $\mu_I$ is the interaction range;
$V_0(J,T)$ is the strength of the interaction, which depends on the total angular momentum $J$ and isospin $T$;
$R_0$ is the radius of the one-body Woods-Saxon potential;  and $\hat{r} = \bbox{r}/r$.

The contact term represented by the Dirac delta function in Eq.~(\ref{SGI_inter})
generates unwanted divergences in momentum space analogous to those present for zero-range interactions.
To rectify this problem, we replace the radial form factors of the multipole expansion of SGI by separable terms, chosen independently of $\ell$ for simplicity. With this choice, the modified interaction MSGI reads:
\begin{eqnarray}
&&V_{J,T}^{MSGI} (\bbox{r_1},\bbox{r_2}) \nonumber \\ &=& V_0(J,T)  \exp \left[ - \left( \frac{r_1 - R_0}{\mu_I} \right)^2 \right] \exp \left[ - \left( \frac{r_2 - R_0}{\mu_I} \right)^2 \right] \nonumber \\ 
                                                &\times&  F(R_0,r_1) F(R_0,r_2) \sum_{\ell=0}^{ \ell_{\rm max}} \bbox{Y}_\ell(\hat{r}_1) \cdot \bbox{Y}_\ell(\hat{r}_2)
\label{MSGI_inter},
\end{eqnarray}
where $\displaystyle F(R_0,r) = \left[ 1 + \exp \left( \frac{r - 2 R_0 - r_F}{\mu_F} \right)\right]^{-1}$ (with $r_F = 1$\,fm and $\mu_F = 0.05$\,fm)
is a Fermi function which makes MSGI practically vanish at $r >2 R_0$.

The surface character of MSGI is incorporated through the Gaussians centered at $R_0$.
Due to the separability of the radial form factors and the presence of the radial Fermi cut-off, two-body radial matrix elements of MSGI are products of one-dimensional integrals which are non-zero only for $0<r<2 R_0$; hence, 
they are as easy to calculate as the  radial integrals of SGI \cite{Mic04}.
The range of MSGI is fixed at $\mu_I=1$\,fm. The 
coupling constants $V_0(J,T)$  are  adjusted to
the  binding energies ground state (g.s.)  and first 2$^+$ state of $^6$He and $^6$Be.
It is important to point out that  the two-body nuclear GSM interaction of Eq.~(\ref{MSGI_inter}) is isoscalar  by construction. That is, in our work, we do not address the  question of  INC  nuclear forces.

The valence space for neutrons and protons consists of all partial waves of angular momentum $\ell=$ 0, 1, and 2. Consequently,
the orbital angular momentum cut-off in Eq.~(\ref{MSGI_inter}) is $\ell_{\rm max}$=2.
The
$p_{3/2}$ wave functions include a $0p_{3/2}$ resonant state and $p_{3/2}$ non-resonant scattering states along a complex contour enclosing the $0p_{3/2}$ resonance in the complex $k$-plane. For the remaining  partial waves,
i.e., $s_{1/2}$, $p_{1/2}$, $d_{3/2}$, and $d_{5/2}$, we take the non-resonant contour  along the real-$k$ axis (the broad $0p_{1/2}$ resonant state plays a negligible role in the g.s.~wave function of $^{6}$He and $^{6}$Be). 
For all contours, the maximal momentum value is $k_{\rm max}=4$\,fm$^{-1}$. The contours have been discretized with up to 80 points.

In calculations for systems having valence protons, one has to consider explicitly the Coulomb interaction. For 
$^5$Li, it is represented by a one-body Coulomb potential of $^4$He. In principle, one could approximate it by a Coulomb potential of a uniformly charged sphere of radius
$R_0$. However, such a potential is inconvenient to use because of its non-analytic
behavior  at $R_0$. Therefore, we use the
dilatation-analytic form of the Coulomb potential $U^{(Z)}_c$ \cite{Sai77,Myo98,idb08}, generated by a Gaussian proton density: 
\begin{eqnarray}
U^{(Z)}_c(r) &=& Z e^2~\frac{{\rm erf}(r/\nu_c)}{r}.
\label{coul}
\end{eqnarray}
In the above equation, $\nu_c=4R_0/(3\sqrt{\pi})$, where $R_0$ is the radius of the WS potential, and $Z$ is the number of protons of the target, e.g., $Z$=2 for the ``proton + $^4$He core" system. 
The above choice of $R_0$ assures that the Coulomb potential given by Eq.~(\ref{coul}) and the uniformly charged sphere potential are equal at $r$=0.

The nucleus $^6$Be has two valence protons outside the $^4$He core. Consequently,  the two-body Coulomb interaction $V_c$ has to be considered.
Unfortunately, the calculation of two-body matrix elements
of $V_c$ in a basis generated by the one-body part of the GSM Hamiltonian is impractical because of difficulties associated with computing two-dimensional integrals with the complex scaling method for  resonant and scattering basis states. A more practical procedure can be developed if one notices that at large distances the Coulomb term $U^{(2)}_c + V_c$ must behave as $\displaystyle U^{(3)}_c(r) \sim 3~e^2~r^{-1}$. 
Consequently, since $U^{(Z)}_c$ is additive in $Z$, one can rewrite the Coulomb interaction  in the $^6$Be Hamiltonian  as $U^{(3)}_c + (V_c - U^{(1)}_c)$.
The short-range character of the operator $V_c - U^{(1)}_c$ suggests using the method which consists of expanding two-body operators in a truncated basis of harmonic oscillator (HO) states~\cite{Gaute_Morten}:
\begin{eqnarray}
V_c - U^{(1)}_c &\simeq& [V_c - U^{(1)}_c]^{(N)} \nonumber \\
&=& P_N (V_c - U^{(1)}_c) P_N \nonumber \\
&=& \sum_{\alpha \beta \gamma \delta}^{N} | \alpha \beta \rangle \langle \alpha \beta | V_c - U^{(1)}_c | \gamma \delta \rangle \langle \gamma \delta |
\label{HO_expansion}
\end{eqnarray}
where Greek letters label HO states, $N$ is the number of HO states used in a given partial wave, and $P_N$ is a projector:
\begin{eqnarray}
P_N = \sum_{\alpha \beta}^{N} | \alpha \beta \rangle \langle \alpha \beta | .
\label{PN_projector}
\end{eqnarray}

To justify the approximation stated in Eq.~(\ref{HO_expansion}), let us consider a normalizable two-body eigenstate $| \Psi \rangle$. $| \Psi \rangle$ can be either bound or resonant
because resonant states become integrable when complex scaling is applied to radial coordinates \cite{Gya71}. 
In this case, $| \Psi \rangle$ can be expanded in the HO basis used in Eq.~(\ref{HO_expansion}). 
According to Eq.~(\ref{PN_projector}), $P_N | \Psi \rangle \rightarrow | \Psi \rangle$ when $N \rightarrow +\infty$. 
Hence, the matrix elements of the operator $[V_c - U^{(1)}_c]^{(N)}$ involving two-body normalizable states converge to those of $V_c - U^{(1)}_c$ when 
$N \rightarrow +\infty$, i.e.~$\displaystyle \langle \Psi_f | [V_c - U^{(1)}_c]^{(N)} | \Psi_i \rangle  \rightarrow \langle \Psi_f | V_c - U^{(1)}_c | \Psi_i \rangle $.
The latter equality is independent of the basis used to expand $| \Psi_i \rangle$ and $| \Psi_f \rangle$. In particular, one can use the Berggren basis for this purpose.
The short-range character of the operator $V_c - U^{(1)}_c$ implies that $\displaystyle \langle \Psi_f | [V_c - U^{(1)}_c]^{(N)} | \Psi_i \rangle $ should converge rapidly with $N$.
This argument can be easily generalized for many-body wave functions 
with more than two particles.

The matrix elements in Eq.~(\ref{HO_expansion}) can be  calculated efficiently using the Brody-Moshinsky transformation. The computation of  one-body overlap integrals  between Berggren basis and  HO states is straightforward, as these  always converge  along the real axis due to the Gaussian tail of HO states; hence,
no complex scaling is needed. The recoil term in Eq.~(\ref{GSMCOSM}) can be treated in the same way as the Coulomb interaction, i.e., by expanding 
$\bbox{p}_i$ in a HO basis \cite{Gaute_Morten}.
The attained precision of calculations on energies and widths is better than 0.2 keV for calculations without recoil and Coulomb terms, and it is around 1\,keV for the full GSM scheme. 

It has to be noted that because our model involves a core, our treatment of the Coulomb interaction is not exact. In particular, we neglect the contribution to the exchange term arising from the core protons. We also ignore other known charge-symmetry breaking electromagnetic terms such as the Coulomb spin-orbit interaction.

\section{Spectroscopic factors in isobaric analog states}\label{spectfac} 

The SF in the GSM framework is given by  the real part of the
squared norm  $S^2$ of the overlap integral between initial and final state in the reaction
channel \cite{Mic07,Mic07a}. The imaginary part of $S^2$, which is an uncertainty of ${\cal R}e$($S^2$), vanishes if  both states in nuclei $A$ and $A$--1 are bound. 
Using a decomposition of the s.p.~channel $(\ell,j)$  in the complete Berggren basis, one obtains:
\begin{equation} S^2 = \int\hspace{-1.4em}\sum_{\mathcal{B}} \langle
\widetilde{\Psi^{J_{A}}_{A}} || a^+_{\ell j}(\mathcal{B}) || \Psi^{J_{A-1}}_{A-1} \rangle^2 
\label{eq3} 
\end{equation}
where $a^+_{\ell j} (\mathcal{B})$ is a creation operator associated with a s.p.~basis state 
$|u_\mathcal{B}\rangle$
and the tilde symbol above bra vectors
signifies that the complex conjugation arising in the
dual space affects only the angular part and leaves the radial
part unchanged.
Since Eq.~(\ref{eq3}) involves summation over all discrete Gamow states and integration over all scattering states along  the complex contour, the final result is independent of  the s.p.~basis assumed. 
This feature is crucial for loosely bound states and near-threshold resonances, where the coupling to the non-resonant continuum can no longer be neglected. 
Indeed, the contribution of the scattering continuum to SFs can be as large as 25\% in such cases \cite{Mic07,Mic07a}. 

In the context of this study,  the direct use of Eq.~(\ref{eq3}) is impractical when assessing effects related to the configuration mixing. Indeed, because of the presence of reduced matrix elements, $S^2 \neq 1$ in the absence of many-body correlations, and its value depends on $j$, $J_{A-1}$, and $J_A$. Hence, we choose to renormalize $S^2$ by dividing it by the extreme single-particle value (obtained by neglecting  two-body interactions). Within this convention, $S^2 = 1$ if configuration mixing is absent.

The SFs for the  two-neutron ($^6$He) and two-proton ($^6$Be) 
g.s. configurations considered in our work correspond to the
$[{^5}{\rm He}({\rm g.s.})\otimes \nu p_{3/2}]^{0^+}$ and
$[{^5}{\rm Li}({\rm g.s.})\otimes \pi p_{3/2}]^{0^+}$   channels, respectively.
For the $T=1$ IASs  in $^6$Li, we consider two channels: 
$[{^5}{\rm He}({\rm g.s.})\otimes \pi p_{3/2}]^{0^+}$ and $[{^5}{\rm Li}({\rm g.s.})\otimes \nu p_{3/2}]^{0^+}$.

\subsection{Stability of HO expansion}

\begin{figure}[hbt]
\begin{center}
\includegraphics[width=0.4\textwidth,angle=00]{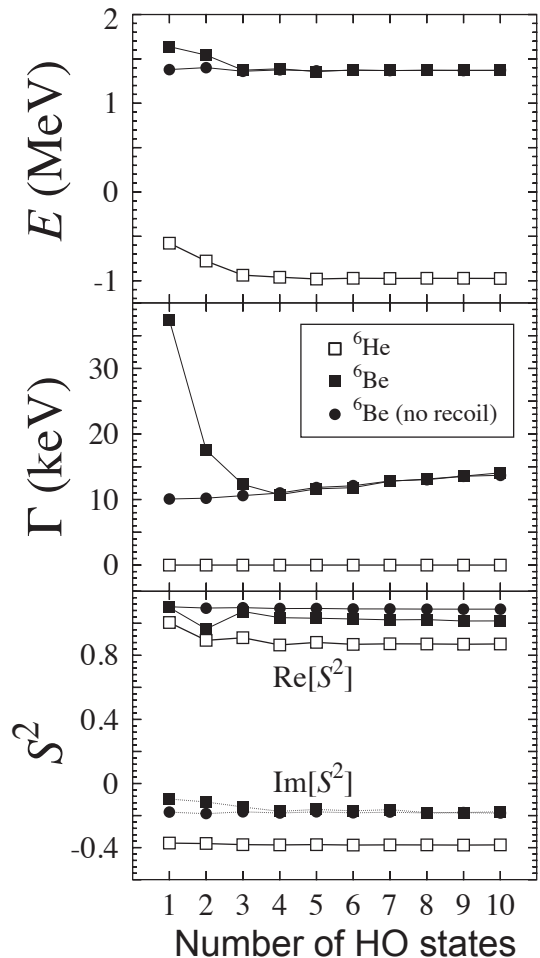}
\caption{The numerical check of the HO expansion (\ref{HO_expansion})
for the total energy (top), width (middle), and spectroscopic factor (bottom)
for g.s.~configurations of $^6$Be (filled symbols) and $^6$He (open squares).
The GSM predictions are plotted as a number of HO states used
in the expansion. The HO length is $b$=2\,fm \cite{Gaute_Morten}.} 
\label{HOcheck}
\end{center}
\end{figure}

The quality of the HO expansion of Eq.~(\ref{HO_expansion})  has been numerically checked for both the Coulomb interaction and recoil term. Figure~\ref{HOcheck}
displays the convergence with respect to the number of HO states used in the expansion for the total energy, width, and SF (real and imaginary part)  
of g.s.~configurations in $^6$Be  and $^6$He. For  $^6$Be, the  results obtained by assuming the inert core (no recoil, i.e., $A_c = +\infty$ in Eq.~(\ref{GSMCOSM})) are also presented. 
It is seen  that with nine HO states per partial wave, one obtains excellent  convergence for both energies   and  wave functions, the latter being represented by SFs.
For the complex energy, the associated numerical error is of the order of
2 keV, and this is well below other theoretical uncertainties of the model.

\subsection{Threshold dependence of spectroscopic factors}

\begin{figure}[hbt]
\begin{center}
\includegraphics[width=7cm,angle=00]{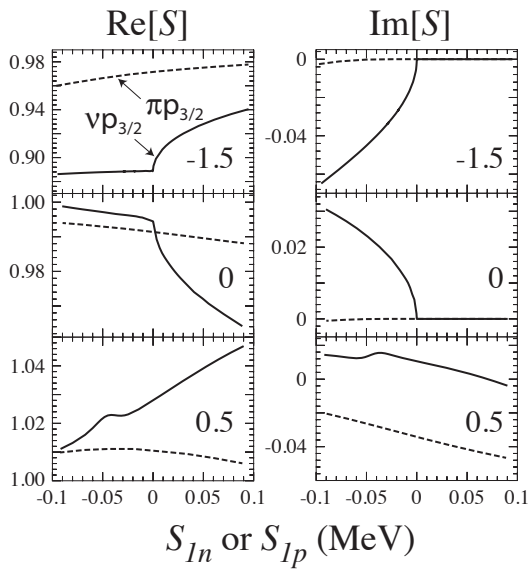}
\caption{The spectroscopic factor $S = \sqrt{S^2}$, i.e., the real and imaginary parts of the square root of the overlap integral (\ref{eq3}) corresponding to
$\langle {^6}{\rm He}({\rm g.s.}) |[{^5}{\rm He}({\rm g.s.})\otimes \nu p_{3/2}]^{0^+}\rangle$ (solid line) and 
$\langle {^6}{\rm Be}({\rm g.s.}) |[{^5}{\rm Li}({\rm g.s.})\otimes \pi p_{3/2}]^{0^+}\rangle$  (dashed line) 
as a function of one-nucleon separation energy ($S_{1n}$ for $^6$He and $S_{1p}$ for $^6$Be)
for three different values of one-particle threshold energy $E_T = -1.5$, 0, and 0.5 MeV  in $A$=5 systems (indicated at the right of each panel).} 
\label{SF_He_Be_COSM}
\end{center}
\end{figure}

The GSM SFs for $^6$He and $^6$Be
are shown in the left column of Fig.~\ref{SF_He_Be_COSM}. The results are plotted as a function of one-nucleon separation energy ($S_{1n}$ for $^6$He and $S_{1p}$ for $^6$Be)
for three different values of  the one-particle threshold energy $E_T$ (i.e., negative of
one-nucleon separation energy) in one-nucleon systems: $^5$He and $^5$Li.

For the bound $A$=5 systems ($E_T=-1.5$ MeV), the SFs in $^6$He and $^6$Be are different in the whole range  of separation energies considered. The difference of the SFs reach the maximum at the one-nucleon emission threshold.
As the separation energy increases (both nuclei become more particle-bound) both SFs slowly approach the value of one, as expected from simple shell-model considerations \cite{Mic07}.
A  characteristic irregularity  in the $\ell=1$ neutron SF  at the neutron emission threshold of $^6$He is the Wigner cusp. 
The cusp is absent in the mirror system $^6$Be as a result of the different asymptotic behavior of the proton  wave function  \cite{Wigner}. 

The energy dependence of SFs changes if the $A$=5 system happens to be  at the particle emission  threshold ($E_T$=0) or is unbound  ($E_T$=0.5 MeV). 
In both situations, a significant difference of SFs in mirror states is seen  in particle stable (positive $S_{1n}$ or $S_{1p}$) $A$=6  systems. 
One may also notice that the Wigner cusp  disappears altogether if $^5$He becomes unbound (cf. $E_T$=0.5\,MeV variant in Fig.~\ref{SF_He_Be_COSM}). 
It is interesting to notice that  SFs can be greater than 1 if the state of the $A-1$ system is particle unstable. 
This unusual situation (see, e.g., ${\cal R}e$($S$) in $^6$He for $E_T=+0.5$ MeV) is discussed below.

The imaginary part  of the expectation value of an operator in a resonant state can be interpreted as the uncertainty in the
determination of this expectation value due to the possibility
of decay during the measuring process \cite{Ber96,Civ99,Hat08,Hat09}.
Figure~\ref{SF_He_Be_COSM} (right column)  shows the uncertainty ${\cal I}m$($S$) of SFs displayed in Fig.~\ref{SF_He_Be_COSM} (left column). 
The uncertainty  vanishes if the wave functions in both $A$=6 and $A$=5 systems are bound with respect to the particle emission. 
Note that in Fig.~\ref{SF_He_Be_COSM}, the appearance of ${\cal R}e$($S$)$>1$ cannot be fully explained by the ${\cal I}m$($S$) plot. 
Indeed, for $^6$He at $E_T$=0.5\,MeV and $0.05 < S_{1n} < 0.1$ MeV,  ${\cal R}e$($S$)$>1$ corresponds to $\vert {\cal I}m$($S)\vert \simeq 0$.

\begin{figure}[hbt]
\begin{center}
\includegraphics[width=7cm,angle=00]{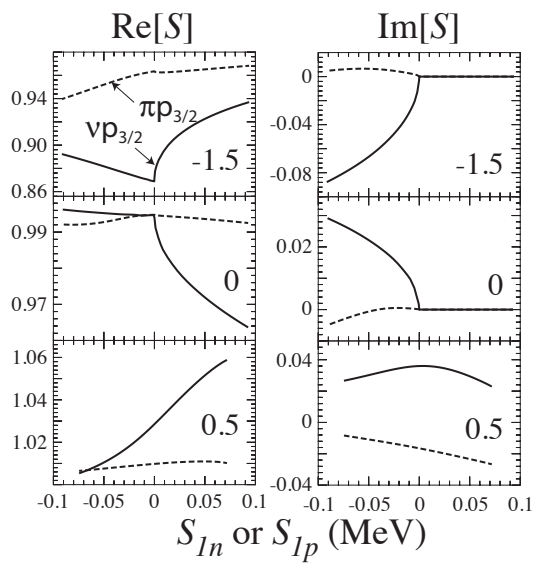}
\caption{Same as in Fig.~\ref{SF_He_Be_COSM} except 
for $\langle {^6}{\rm Li}(T=1) |[{^5}{\rm Li}({\rm g.s.})\otimes \nu p_{3/2}]^{0^+}\rangle$ (solid line) and 
$\langle {^6}{\rm Li}(T=1) |[{^5}{\rm He}({\rm g.s.})\otimes \pi p_{3/2}]^{0^+}\rangle$  (dashed line) 
as a function of one-nucleon separation energy ($S_{1n}$ for the former and $S_{1p}$ for the latter) 
for three different values of one-particle threshold energy $E_T = -1.5$, 0, and 0.5 MeV  in $A$=5 systems.
The proton and neutron threshold energies are assumed to be identical.} 
\label{Re_SF_Li_COSM}
\end{center}
\end{figure}

For the  $J^\pi=0^+$ IAS  in $^6$Li,
we consider two different SFs for the $(\ell,j) = p_{3/2}$ channel,
associated with adding a proton to $^5$He  or a neutron to $^5$Li (see Fig.~\ref{Re_SF_Li_COSM}).
They are plotted in Fig.~\ref{Re_SF_Li_COSM}
as a function of one-proton (or one-neutron) separation energy.
The channel wave functions
$|[{^5}{\rm He}({\rm g.s.})\otimes \pi p_{3/2}]^{0^+}\rangle$
and $|[{^5}{\rm Li}({\rm g.s.})\otimes \nu p_{3/2}]^{0^+}\rangle$
are obviously not orthogonal, as they both share the dominant
$|[{^4}{\rm He}({\rm g.s.})\otimes \pi~0p_{3/2} \otimes \nu~0p_{3/2}]^{0^+}\rangle$ component. The two considered SFs for 
$^6$Li  differ only by  continuum couplings induced in the proton and neutron channels.
Comparing Fig.~\ref{SF_He_Be_COSM} and Fig.~\ref{Re_SF_Li_COSM}, one can see 
$p_{3/2}$ proton and neutron SFs factors are very similar in both cases. Still, slight differences are present. For instance,  at $E_T = 0.5$ MeV, 
small irregularities seen in   $^6$He SFs are absent in the neutron SF for  $^6$Li. A close inspection of proton SFs for 
 $^6$Li reveals the presence of threshold cusps at zero separation, absent on the
 $^6$Be case.  
This effect  can be explained in terms of the channel
coupling, or flux conservation  \cite{Mic07,Mic07a}.
Indeed, since in our model calculations
both proton and neutron channels open at threshold energy, the coupling between
proton and neutron channels can generate non-analyticities in proton SFs, even though Wigner estimates for proton cross sections are analytical at the threshold energy.
 
To study the sensitivity of results to the CM treatment, we carried out a set of calculations assuming the inert core (no recoil). 
The results are practically identical to those of Figs.~\ref{SF_He_Be_COSM} and \ref{Re_SF_Li_COSM}. 
The only noticeable difference is the absence of a small fluctuation at $S_{1n}$$\approx$--0.05\,MeV seen in the real and imaginary parts of SF for $^6$He.

\begin{figure}[hbt]
\begin{center}
\includegraphics[width=8cm,angle=00]{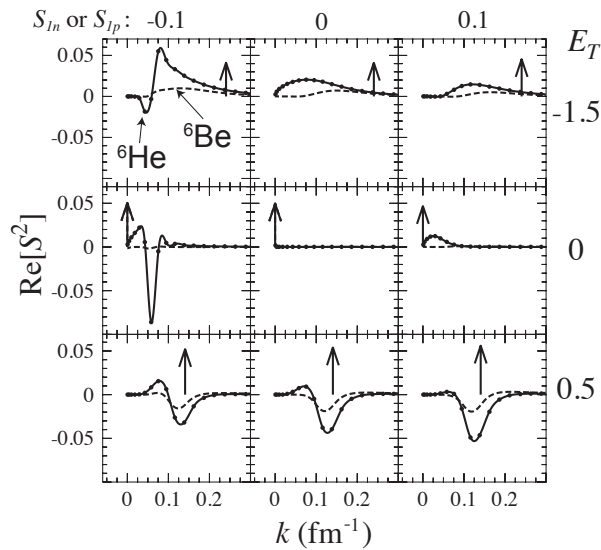}
\caption{The distribution of the real part of ${\cal R}e$($S_{\rm c}^2$) in $^6$He (solid line) and $^6$Be (dashed line) with respect to the $3/2^-$ scattering states of the $^5$He and $^5$Li systems,
ordered according to their real-$k$ value. The calculations were performed for
 three different values of one-particle threshold energy $E_T$=$-1.5$\,MeV (top), 0 (middle), and 0.5\,MeV (bottom) in $A$=5 systems.
The arrows indicate contributions from resonant states. In order to  
facilitate presentation, the pole contributions were multiplied by a scaling factor of 0.05.} 
\label{Re_SF_3x3_COSM}
\end{center}
\end{figure}

\begin{figure}[hbt]
\begin{center}
\includegraphics[width=8cm,angle=00]{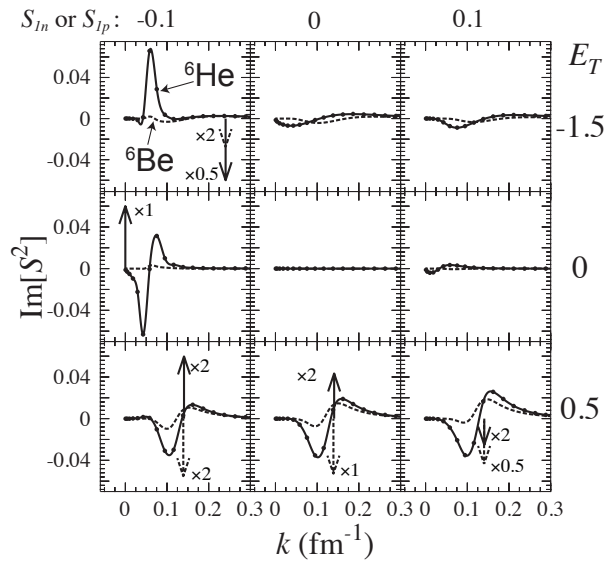}
\caption{Same as in Fig.~\ref{Re_SF_3x3_COSM}, except for the imaginary part of  ${\cal I}m$($S_{\rm c}^2$).
The arrows indicate contributions from $0p_{3/2}$ resonant states if those  
are not negligible. In order to facilitate presentation, the pole  
contributions were multiplied by the scaling factors indicated in the figure.}
\label{Im_SF_3x3_COSM}
\end{center}
\end{figure}
Another reason for the occurrence of ${\cal R}e$($S$)$>1$ in some cases, is the interplay between a  final state $\Psi^{J_{A-1}}_{A-1}\equiv\Psi^{J_{A-1}}_{A-1;{\rm R}}$ (the many-body resonance) 
and states of the non-resonant scattering continuum  $\{ \Psi^{J_{A-1}}_{A-1;{\rm c}} \}$ with energies close to the resonance energy \cite{Ber96}.
The contributions to SFs $S_{\rm c}$ in $^6$He and $^6$Be coming from  the non-resonant continuum
are shown in Figs.~\ref{Re_SF_3x3_COSM} and \ref{Im_SF_3x3_COSM} together with contributions from resonant states.

In all situations, the contribution of the Gamow resonance to ${\cal R}e$($S^2$) is dominant. It is interesting to note that
 the impact of the non-resonant continuum does depend on
$E_T$ and $S_{1p}/S_{1n}$. For $E_T \leq 0$ and $S_{1p}/S_{1n} \geq 0$, i.e., for $A$=5 and $A$=6 bound ground states, the non-resonant continuum contribution is basically negligible.
This is also the case for $^6$Be when $E_T \leq 0$ and $S_{1p}/S_{1n} < 0$, i.e.,  for a bound $^5$Li but an unbound $^6$Be.
However, when either the $A$=5 g.s.~is unbound ($E_T > 0$) or $E_T \leq 0$ and $S_{1p}/S_{1n} < 0$ for $^6$He, i.e., when $^6$He is unbound with respect to a bound $^5$He,
the non-resonant continuum plays a significant role in both real and imaginary parts. 
In particular, when $E_T > 0$ and $S_{1p}/S_{1n} \geq 0$, the contribution from the non-resonant continuum  ${\cal R}e$($S_{\rm c}^2$) becomes negative, which translates into a value of ${\cal R}e$($S^2$) that exceeds one.
One can also see that ${\cal I}m$($S_{\rm c}^2$) is comparable to ${\cal R}e$($S_{\rm c}^2$), even though this does not occur every time ${\cal R}e$($S^2$)$>1$. The lesson learned from this discussion is that 
 the SF obtained by considering the many-body resonance only may often be a poor approximation to the total SF, which can contain appreciable  non-resonant contributions.

\section{Isospin mixing in $^6$He, $^6$Li, and $^6$Be} \label{HeLiBe}

So far, we have discussed prototypical $T$=1 multiplet
`$^6$He',`$^6$Li', and `$^6$Be' with equal proton and neutron separation energies 
to study the effect of different asymptotic behavior on the configuration mixing in the vicinity of one-nucleon thresholds. 
In a realistic situation, however, particle emission thresholds change within the isotriplet due to the Coulomb interaction. To assess this effect,
we shall now apply the GSM to describe spectra and SFs for the $0^+$ ground states 
and  first excited $2^+$ states of $^6$He and $^6$Be, and the IASs in $^6$Li. 
In calculations involving $^5$He and $^6$He, we use 
the ``$^{5}$He" WS parameter set and the MSGI interaction with the strengths: $V_0(J=0,T=1)=-15.193$ MeV$\cdot{\rm fm}^3$, $V_0(J=2,T=1)=-12.505$ MeV$\cdot{\rm fm}^3$.
For $^6$He,
this Hamiltonian yields:  $E_{0^+}=-0.974$ MeV,  $E_{2^+}=+0.823$ MeV, and $\Gamma_{2^+}=+89$ keV. The experimental values are very close:
$E_{0^+}^{({\rm exp})}=-0.973$ MeV, $E_{2^+}^{({\rm exp})}=+0.824$ MeV, and
$\Gamma_{2^+}^{({\rm exp})}=+113$ keV. All binding energies are given relative to the binding energy of the $^4$He core.

\begin{table*}[htb]\label{0+_amplitudes}
\renewcommand{\arraystretch}{1.2}
\renewcommand\tabcolsep{3pt}
\caption[T3]{Squared GSM amplitudes of 
the $J^\pi$=0$^+$ IASs of the isotriplet
$^6$He, $^6$Li, and $^6$Be. The symbols $S1$ and $S2$ indicate 
configurations with one and two particles in the non-resonant continuum, respectively. The results for $^6$He assuming a rigid $^4$He core are shown in the third column.}
\label{table1}
\begin{ruledtabular}
\begin{tabular}{l|llllll} 
$(C_k)^2$          & ~~~$^6$He       & $^6$He (rig.core)  & ~~$^6$Be (V1)         & ~~$^6$Be (V2)       & ~~$^6$Li (V1)       & ~~$^6$Li (V2)  \\ 
\hline
$(0p_{3/2})^2$      & 0.750$-$i0.692  & 0.798$-$i0.732    & 1.090$-$i0.243        & 1.107$-$i0.288      & 0.994$-$i0.587      & 0.949$-$0.614
\\
$(S1)_{\pi p_{3/2}}$   & ~~~~~~---       & ~~~~~~---         & $-$0.115+i0.218       &$-$0.143+i0.255      & $-$0.084+i0.226     & $-$0.050+i0.244
\\
$(S1)_{\nu p_{3/2}}$   & 0.243+i0.619    & 0.244+i0.668      & ~~~~~~---             & ~~~~~~---           & 0.066+i0.308        & 0.0797+i0.314
\\
$(S2)_{s_{1/2}}$     & 0.009+i0.0      & 0.0+i0.0          & 0.022+i0.0            & 0.023+i0.004        & 0.011+i0.0          & 0.010+i0.0
\\
$(S2)_{p_{1/2}}$     & 0.012+i0.0      & 0.013+i0.0        & 0.008+i0.001          & 0.009$-$i0.0        & 0.011+i0.0          & 0.012+i0.0
\\
$(S2)_{p_{3/2}}$     & $-$0.049+i0.074 & $-$0.063+i0.065   & $-$0.030+i0.029       & $-$0.028+i0.034     & $-$0.033+i0.054     & $-$0.033+i0.055  
\\
$(S2)_{d_{3/2}}$     & 0.002+i0.0      & 0.001+i0.0        & 0.002+i0.0            & 0.002$-$i0.0        & 0.002+i0.0          &  0.002+i0.0
\\
$(S2)_{d_{5/2}}$     & 0.032+i0.0      & 0.006+i0.0        & 0.025$-$i0.0          & 0.031$-$i0.04       & 0.031+i0.0          & 0.031+i0.0 \\
\end{tabular} 
\end{ruledtabular}
\end{table*}
\begin{table*}[htb]\label{2+_amplitudes}
\renewcommand{\arraystretch}{1.2}
\renewcommand\tabcolsep{3pt}
\caption[T3]{Same as in Table~\ref{table1}, except for the first  $2^+$ state.}
\label{table2}
\begin{ruledtabular}
\begin{tabular}{l|llllll} 
$(C_k)^2$                       & ~~~$^6$He        & $^6$He (rig.core) & ~~$^6$Be (V1)       & ~~$^6$Be (V2)     & ~~$^6$Li (V1)     & ~~$^6$Li (V2)  \\ 
\hline
$(0p_{3/2})^2$                   & 1.132+i0.006    & 1.149$-$i0.022    & 0.977$-$i0.023      &  0.987$-$i0.0267  & 1.036$-$i0.024    & 1.049$-$i0.023
\\
$(S1)_{\pi p_{1/2}}$               & ~~~~~~---       & ~~~~~~---         & $-$0.004$-$i0.001   &  $-$0.001+i0.001  & 0.0$-$i0.0        & 0.0$-$i0.0
\\
$(S1)_{\pi p_{3/2}}$               & ~~~~~~---       & ~~~~~~---         & $-$0.003+i0.022     &  $-$0.011+i0.027  &$-$0.001+i0.001    & $-$0.007+i0.0
\\
$(S1)_{\nu p_{1/2}}$               & 0.0$-$i0.001    &  0.003$-$i0.002   & ~~~~~~---           & ~~~~~~---         & 0.0$-$i0.0        & 0.0$-$i0.0
\\
$(S1)_{\nu p_{3/2}}$               & -0.142$-$i0.009 &  $-$0.147+i0.016  & ~~~~~~---           & ~~~~~~---         & $-$0.0492+i0.021  & $-$0.056+i0.019
\\
$(S2)_{\pi p_{1/2}~\pi p_{3/2}}$     & ~~~~~~---       & ~~~~~~---         &  0.001$-$i0.0       &  0.001$-$i0.0     & ~~~~~~---         & ~~~~~~---
\\
$(S2)_{\nu p_{1/2}~\nu p_{3/2}}$     & 0.001+i0.0      & 0.001+i0.001      & ~~~~~~---           & ~~~~~~---         & ~~~~~~---         & ~~~~~~---
\\
$(S2)_{p_{3/2}}$                  & $-$0.004+i0.006 & $-$0.005+i0.006   & 0.003+i0.006        & 0.002+i0.003      & $-$0.004+i0.008   & $-$0.004+i0.008
\\
$(S2)_{\pi d_{3/2}~\pi d_{5/2}}$     & ~~~~~~---       &  ~~~~~~---        & 0.001$-$i0.001      & 0.001$-$i0.001    &  ~~~~~~---        & ~~~~~~---
\\
$(S2)_{\nu d_{3/2}~\nu d_{5/2}}$     & 0.001$-$i0.0    & 0.0+i0.0          & ~~~~~~---           & ~~~~~~---         &  ~~~~~~---        & ~~~~~~---
\\
$(S2)_{\pi d_{3/2}~\nu d_{5/2}}$     & ~~~~~~---       &  ~~~~~~---        & ~~~~~~---           & ~~~~~~---         & 0.001$-$i0.0      & 0.001$-$i0.0
\\
$(S2)_{\nu d_{3/2}~\pi d_{5/2}}$     & ~~~~~~---       &  ~~~~~~---        & ~~~~~~---           & ~~~~~~---         & 0.001$-$i0.0      & 0.001$-$i0.0
\\
$(S2)_{d_{5/2}}$                  & 0.010$-$i0.002  & 0.0+i0.0          & 0.022$-$i0.006      &  0.020$-$i0.007   & 0.015$-$i0.005    & 0.014$-$i0.003 \\
\end{tabular} 
\end{ruledtabular}
\end{table*}

In the case of $^6$Li and $^6$Be, we carry out calculations in two variants. In variant V1, we take the same WS potential as for the He isotopes. 
Here, isospin is explicitly broken by the one-body Coulomb potential and the two-body Coulomb interaction between valence protons. In variant V2,
the depth of the WS potential has been changed to 47.563\,MeV, in order to obtain an overall agreement for the binding  energies and widths of $3/2_1^-$ and $1/2_1^-$ resonances in  $^5$Li and the  $0^+$ g.s. of
$^6$Be. The readjustment of the one-body potential in V2 is supposed to account for the impact of the missing Coulomb terms, see discussion at the end of Sec.~\ref{model}. 
In both variants, MSGI strengths are the same as in the  He calculation.

The predicted  g.s. energy of $^6$Be,   ($E_{0^+}$=1.653\,MeV,  $\Gamma_{0^+}$=41\,keV) in V1 and ($E_{0^+}$=1.371\,MeV, $\Gamma_{0^+}$=14\,keV) in V2, is close to experiment: 
($E_{0^+}^{({\rm exp})}$=1.371\,MeV, $\Gamma_{0^+}^{({\rm exp})}$=92\,keV).
For the first ${2^+}$ state, we obtain:
($E_{2^+}$=2.887\,MeV,  $\Gamma_{2^+}$=0.986\,MeV) in V1 and
($E_{2^+}$=2.679\,MeV,  $\Gamma_{2^+}$=0.804\,MeV) in V2. The experimental energy is  ($E_{2^+}^{({\rm exp})}$=3.041\,MeV, $\Gamma_{2^+}^{({\rm exp})}$=1.16\,MeV). 

Turning to the $0^+$  IAS of $^6$Li, the predicted energy is
($E_{0^+}$=0.0866\,MeV, $\Gamma_{0^+}$=8.85$\cdot$10$^{-3}$\,keV) in V1 and ($E_{0^+}$=-0.0706\,MeV, $\Gamma_{0^+}$=9.13$\cdot$10$^{-3}$\,keV) in V2. This is fairly close to  the experimental value 
($E_{0^+}^{({\rm exp})}$=-0.136\,MeV, $\Gamma_{0^+}^{({\rm exp})}$=8.2\,eV).
For the  $2^+$ IAS in $^6$Li, we obtain  
($E_{2^+}$=1.667\,MeV,  $\Gamma_{2^+}$=0.404\,MeV) in V1 and
($E_{2^+}$=1.569\,MeV,  $\Gamma_{2^+}$=0.329\,MeV) in V2. Both variants are in a very reasonable agreement with experimental energy: $E_{2^+}^{({\rm exp})}$=1.667\,MeV, $\Gamma_{2^+}^{({\rm exp})}$=0.541\,MeV.

The corresponding g.s.~SFs for $^6$He and $^6$Be (in V1) are 
$S^2$=0.87$-$i0.383 and 1.015$-$i0.147, respectively, while for $^6$Li, they are 1.061$-$i0.280  for $\pi p_{3/2}$   and 0.911$-$i0.361 for $\nu p_{3/2}$. 
For the $2_1^+$  state, the SFs are $S^2$=1.061+i0.0011 for $^6$He, 0.973$-$i0.0142 for $^6$Be,
0.987$-$i3.26$\cdot$10$^{-3}$ for $^6$Li ($\pi p_{3/2}$), and 1.034$-$i0.0235 for the $^6$Li ($\nu p_{3/2}$).
In spite of the fact that both real and imaginary energies in V1 and V2 are slightly different, 
the SFs for $^6$Be in V2 are very close to those obtained in V1. Namely,
$S^2$=1.015$-$i0.177 for the g.s.~and
$S^2$=0.978$-$i0.016 for the $2_1^+$ state.
The V2 values of  SFs in  the 0$+$ state of $^6$Li are
$S^2$=1.028$-$i0.300 ($\pi p_{3/2}$) and $S^2$=0.898$-$i0.369 ($\nu p_{3/2}$),
while for the $T$=1 2$^+$ state they are:
$S^2$=0.993$-$i3.190$\cdot$10$^{-3}$ ($\pi p_{3/2}$)  and $S^2$=1.043$-$i0.0224.
This can be seen from Table~\ref{table1} by comparing the corresponding GSM wave function amplitudes for $^6$Be (columns 4 and 5) and  $^6$Li (columns 6 and 7). 

While their mean values differ by about 15\%,
considering large imaginary parts,  the SFs  predicted  
for the $J^\pi$=0$^+$ IASs of the isotriplet, agree within  calculated uncertainty. 
However, by examining the GSM wave function amplitudes displayed in Table~\ref{table1},
one notes that SFs, being  integrated measures,  do not tell the whole story. The main effect of the Coulomb interaction is the change in distribution of the
$(0p_{3/2})^2$ and
$(S1)_{p_{3/2}}$ g.s.~components, the latter involving one particle in the
non-resonant $p_{3/2}$ continuum.  As a result, a rather different interference pattern between the resonant $0p_{3/2}$ state and the non-resonant continuum is predicted for $^6$He and $^6$Be,
and between resonant and non-resonant states of a different type (proton or neutron) in  $^6$Li.

For the $2^+$  IASs, a meaningful comparison of SFs can be done as they have small imaginary parts.  This is a consequence of the smaller configuration mixing induced by the nuclear interaction. Indeed, as shown in Table~\ref{table2}, the structure of $2^+$ 
states is dominated by the resonant $(0p_{3/2})^2$ component. 
Here we conclude that GSM predicts a mirror symmetry-breaking in  SFs of the order of  $5\%$.

To assess the impact of the recoil term on our findings, we carried out 
calculations in which the recoil  of the core is ignored.
In this case, the  coupling constants refitted to the data  are 
$V_0(J=0,T=1)=-18.237$ MeV$\cdot{\rm fm}^3$ and $V_0(J=2,T=1)=-14.942$ MeV$\cdot{\rm fm}^3$, while the depth of the proton WS potential is now 47.5\,MeV.
Without recoil,
energy observables are very similar to those  obtained in  full  calculations. Namely,  for $^6$He,  only the width of the first excited state differs by a few keV, as it becomes $\Gamma_{2^+}=+84$ keV. 
The energy and width of the $^6$Be g.s.~in V2 remain  the same as with recoil, while
there appears a small change for the 
first excited state of $^6$Be: $E_{2^+}=+2.702 $ MeV and $\Gamma_{2^+}=+0.755$ MeV. 
For $^6$Li, the energy of the 0$^+$ state differs by a few keV in V1 and around 20 keV in V2, while the width remains practically unchanged. For the 2$^+$ state
in $^6$Li, changes are of the order of tens of keV.

The changes in
SFs due to recoil are  small as well. To show it explicitly, in Table~\ref{table1} we compare 
the GSM amplitudes  of the ground-state
wave function of  $^6$He  in the COSM variant (second column) and assuming the rigid $^4$He core (third column). 
The main effect of recoil  is to slightly redistribute partial wave occupations,  in particular the $(d_{5/2})^2$ contribution. For instance, for the $^6$He g.s.,
the sum of the square of amplitudes belonging to the $(d_{5/2})^2$ channel is 6$\cdot 10^{-3}$ without recoil, while it is 3.2$\cdot 10^{-2}$ with the full treatment of recoil. 
For  $^6$Be, not shown in Table~\ref{table1}, these numbers in V2 translate to 4.2$\cdot 10^{-3}$ and 3.1$\cdot 10^{-2}$, respectively. There is also a small increase of amplitudes in other continuum channels, e.g., $(s_{1/2})^2$
but those wave function components are very small.

Another way of assessing the degree of isospin mixing is by inspecting the structure of IAS within the isomultiplet. To this end, we carried out calculations for the isotriplet $^6$He, $^6$Li, and $^6$Be in V1+COSM using the common neutron s.p. basis of $^6$He. In this way, the 
isospin operator 
\begin{equation} 
\hat T^- = \int\hspace{-1.4em}\sum_{\mathcal{B}}  a^+_{\ell j \tau_z=-1/2}(\mathcal{B}) a_{\ell j \tau_z=1/2}(\mathcal{B}) 
\label{tminus} 
\end{equation}
is properly defined \cite{[Mil08]}. The  numerical error due to the use of neutron s.p. basis on the ground state energy of $^6$Be is very small: it is about 20\, keV for the real energy 
and 5\, keV for the with, and this accuracy is more than  sufficient for the purpose of our IAS analysis. The isobaric analogs of the $T$=1 states
in $^6$He are given by: 
\begin{subequations} \label{iastates} 
\begin{eqnarray}
|^6{\rm Li, IAS}\rangle &=& \frac{1}{\sqrt{2}} \hat T^- |^6{\rm He}\rangle,
\label{IASLi}\\
|^6{\rm Be, IAS}\rangle &=& \frac{1}{2} (\hat T^-)^2 |^6{\rm He}\rangle. \label{IASBe}
\end{eqnarray}
\end{subequations}
The IAS content of a GSM state  can be obtained by calculating its overlap 
with the state (\ref{iastates}). For the 0$^+$ state of $^6$Li, the 
squared overlap is $\langle ^6{\rm Li}|^6{\rm Li, IAS}\rangle^2$ =0.995.
This  indicates that the lowest $0^+$ state in $^6$Li  is indeed an excellent isobaric analog of $^6$He g.s. Indeed, the  corresponding average
isospin  value  \cite{Mic04}:
\begin{eqnarray}
T_{av} = \frac{-1 + \sqrt{1 + 4 \langle \Psi | \hat{T}^2 | \Psi \rangle}}{2} \label{Tav_def}
\end{eqnarray}
is $T_{av}$=0.9994.

For the ground state of $^6$Be,  
 $\langle ^6{\rm Be}|^6{\rm Be, IAS}\rangle^2$=0.951--i0.050, i.e., the mean
 value of the squared amplitude exhibits a reduction with respect to the perfect isospin invariance. This result is consistent with
the large difference between GSM wave functions of $^6$He and $^6$Be: a significant component of the $^6$Be g.s. wave function corresponds to a non-resonant continuum of $^6$He.   Interestingly, the total isospin of $^6$Be states is perfectly conserved in our GSM space. Indeed, having two valence protons, wave functions of  $^6$Be are completely aligned in isospace, 
{\it regardless} of the strength of Coulomb interaction. The isospin breaking in  $^6$Be can only happen through core polarization effects, i.e.,  core-breaking excitations \cite{[Ber72],[Sat78],[Sat09]}.  Since $^4$He is a very rigid core, one expects a fairly pure isospin in the low-lying states of $^6$Be. 

A similar situation is expected for any isospin-aligned shell-model state corresponding to a semi-magic nucleus having $Z_{\rm val}$ valence protons (or $N_{\rm val}$ neutrons). If one disregards core-breaking effects, such a state has pure isospin $T=Z_{\rm val}/2$ (or $T=N_{\rm val}/2$), in spite of the presence of INC interactions that manifestly break isospin.  
This  is a nice example of a more general phenomenon called partial dynamical symmetry, i.e., a symmetry that is obeyed by a subset of eigenstates, but is not shared by the Hamiltonian \cite{[Lev02],[Lev10]}. We note that 
while $\hat T^2$ and $\hat T_z$ are preserved in the isospin-aligned  states, this is not the case for $\hat T^\pm$ operators connecting $^6$Be with $^6$Li and $^6$Li with $^6$He, that are affected by isospin mixing.

\section{Conclusions}\label{conclusions}

There are several sources of isospin (and mirror) symmetry-breaking in atomic nuclei. Probably the most elusive are consequences of the threshold effect \cite{Wigner}
and the Coulomb-nuclear interference effect \cite{Ehr51,Tho52,Lan58}. The  open quantum system formulation of the GSM makes it possible to address the question of the continuum-induced isospin  symmetry-breaking in a comprehensive and non-perturbative way,
in terms of  the configuration mixing involving bound and unbound states.

As compared to previous GSM studies, present calculations are based on a newly developed finite-range residual interaction MSGI. The Coulomb interaction and recoil term are treated by means of the HO expansion technique. The stability of this expansion  has been numerically checked with a very encouraging result:  with only nine HO states per partial wave, one obtains excellent  convergence for both energies and  wave functions.

To study the sensitivity of results to the CM treatment, we carried out two sets of calculations: one in COSM coordinates in which the core recoil is treated exactly  and another one assuming no recoil. We find that the results of both variants are very close for both energies and SFs; hence, the details of CM  treatment do not impact the conclusions of our work.

We have shown that the energy dependence of  SFs  of mirror nuclei is different.   
Realistic estimates for the isotriplet $^6$He and $^6$Be yield an effect in  SFs of the 2$^+$  state which is in a range of  several percent. This is consistent with results of recent cluster-model studies \cite{Tim08,Tim08a}. 
For the $0^+$ configuration, the situation is different. Here,
the mean values of  SFs differ by about 16\% and a  different interference pattern between the resonant $0p_{3/2}$ components and the non-resonant $p_{3/2}$ continuum is predicted.
However, due to appreciable imaginary parts, hence large 
uncertainty,  g.s.~SFs in $^6$He and $^6$Be, and SFs for the 0$^+$ analog state in  $^6$Li,  calculated in GSM
do not offer a clear measure of the mirror symmetry-breaking.
The behavior of SFs in  $^6$Li follows that predicted for $^6$He and $^6$Be. Interestingly, proton spectroscopic factors show the presence of threshold anomalies due to the strong coupling with the neutron channel.

Due to the partial dynamical isospin symmetry present in the GSM  wave functions of  $^6$Be, the low-lying states in this isospin-stretched ($T$=1, $T_z$=--1) system are expected to show very weak isospin breaking effects. This is in spite of  Coulomb interaction present in this nucleus. For the $T_z$=0 member of the isotriplet, $^6$Li, the isospin symmetry is explicitly broken in the GSM space as a result of  mixing between $T$=0 and $T$=1 states but the resulting mixing is very weak. We thus conclude that the large mirror symmetry breaking effects seen in binding energies and SFs of the isotriplet are related to $\hat T^\pm$ components rather than the total isospin.

In summary, the coupling to the non-resonant continuum can give rise to
isospin  and mirror symmetry-breaking effects that are configuration dependent. 
Explanations of mirror symmetry
breaking based on the traditional close quantum system formulation of
the nuclear shell model sometimes invoke INC
{\em nuclear}  effective interactions \cite{Ben07,Zuk02}. We would like to
point out that any attempt to extract  such  interactions from
spectroscopic data should first account for the coupling to the
many-body continuum in the presence of isospin-conserving nuclear forces. If neglected, or not treated carefully, the
continuum effects can  alter results of such analyses.

This work was supported in part by the Office of Nuclear Physics,  U.S. Department of Energy under
Contract No. DE-FG02-96ER40963 (University of Tennessee), by the
CICYT-IN2P3 cooperation, and by the Academy of Finland and University of
Jyv{\"a}skyl{\"a} within the FIDIPRO programme. WN acknowledges support
from the  Carnegie Trust and the Scottish Universities Physics Alliance
during his stays in Scotland.

\end{document}